\documentstyle[12pt]{article}
\def \be{\begin{equation}}
\def \ee{\end{equation}}
\normalbaselines
\begin{document}
\begin{flushright}
TIFR/TH/96-57\\
18~October~1996
\end{flushright}
\bibliographystyle{unsrt}
\vskip0.5cm
\baselineskip=.8cm
\begin{center}
{\LARGE \bf The $\Delta B =- \Delta Q$ transitions and
$B_d\leftrightarrow \bar B_d$ oscillations}\\ 
[7mm] {\bf G.V. Dass$^a$ and
K.V.L. Sarma}$^{b,*}$\\ 
[3mm] $^a${\it Department of
Physics, Indian Institute of Technology, Powai, Mumbai, 400 076, 
India
}\\
\vskip0.3cm
$^b${\it Tata Institute of Fundamental Research, Homi Bhabha Road,
Mumbai, 400 005, India }
\\
[10mm]
\end{center}
\bigskip
\begin{center}
Abstract
\end{center}
\bigskip
\baselineskip=.8cm
We estimate the product of the relative 
strength of the $\Delta B = -\Delta Q$ amplitude in the decay $B_d^0
\rightarrow D^{*} \ell \bar \nu_{\ell }$ and the width-difference 
parameter $y$. For this we have used the data on time-dependence of
$B_{d}^0\bar B_{d}^0$ oscillations in $Z$ decays and the fraction of
like-sign dilepton events at the $\Upsilon (4S)$.

\bigskip

\bigskip


\vfill
$^*$~E-mail: kvls@theory.tifr.res.in;~~fax: 091 22 215 2110 

\newpage
\baselineskip=.8cm
There is now a considerable body of experimental evidence in favour of
$B_d^0 \bar B_d^0$ oscillations. This comes from the studies of charge
correlations in $Z\rightarrow b\bar b$ decays at the $Z$ resonance
\cite{MIX} and also in hadron collisions wherein the bottom 
flavour is produced \cite{HAD}. The general strategy adopted by the
various experimental groups (see e.g., Ref \cite{WU}) can be briefly
stated as follows: (a) tag the bottom flavour at production (when $Z$
decays) by the charge of a high $p_T$ lepton, or by the ``jet
charge'', (b) in the opposite side hemisphere, look for $B^0$ meson
decay events that contain, say, $D^{*\mp }$ and $\ell ^{\pm }$, (c)
measure the displacement of the decay vertex from the production
vertex, (d) estimate the parent $B^0$ momentum, and (e) determine the
duration of propagation of the neutral beon. From such measurements it
has been shown that the fraction of events arising from $B_d\bar B_d$
mixing has a sinusoidal time-dependence characteristic of
oscillations.

In these analyses, it has been implicitly assumed that the quark model
rule $\Delta B = \Delta Q$ is valid in the semileptonic decays of
neutral $B$ mesons. Here we wish to examine whether any information on
this rule can be gleaned from the oscillation data obtained at the $Z$
resonance. To this end we make use of the mass difference $\Delta m$
extracted from oscillation data and the like-sign dilepton fraction
$\chi _d $ obtained in the decay of $\Upsilon (4S)$ state. We are able
to extract the product of the relative strength $\rho $ of the $\Delta
B = -\Delta Q$ amplitude in the $ D^* \ell \nu $ mode of semileptonic
$B_d^0$ decay and the width difference parameter
\be y =\frac {\Gamma _2 -\Gamma _1}{2\Gamma }~;~~\Gamma = {\Gamma _2
+ \Gamma _1 \over 2}~.   \ee

In regard to the determination of the initial flavour (also known as
production tag) we consider, for illustration, the jet-charge
technique (see, e.g., \cite{OP94}). Consider the jet produced by a $b$
quark coming from $Z$ decay. This jet can show up as having the normal
jet charge ($Q_J=-1/3$) with a probability $b_n$, or an abnormal jet
charge ($Q_J=+1/3$) with a probability $b_a$. The abnormal jet charge
arises in the Standard Model due to the $b$ fragmenting into a $\bar
B_d$ or $\bar B_s$ which undergoes oscillation to a neutral meson with
positive bottom flavour. Therefore the probabilities associated with
jet production are
\be b_n={\rm Prob}(b \rightarrow J^{-1/3}),~~b_a={\rm Prob}(b
\rightarrow J^{+1/3}), \ee
\be \bar b_n={\rm Prob}(\bar b \rightarrow J^{+1/3}),~~\bar b_a={
\rm Prob}(\bar b \rightarrow J^{-1/3}),
\ee
wherein the superscripts denote the jet charges. Requiring $CP$
invariance would imply the relations $b_n=\bar b_n$ and $b_a=\bar
b_a$.

As for the decay tag, we take the time $t=0$ when the primary decay
$Z\rightarrow b\bar b$ (we ignore $Z$ decays involving multiple $b
\bar b$ pairs) occurs. Let the $b$ quark produce a $\bar B_d^0 $ 
meson which undergoes flavour oscillations during its propagation and
decays semileptonically at time $t$. We shall, for definiteness, focus
on the decay mode $B\rightarrow D^*(2010)\ell \nu $. The corresponding
normal and abnormal decay rates are denoted by
\begin{eqnarray}
\bar B_n(t) &\equiv & \Gamma (\bar B^0_d(t) \rightarrow  D^{*+}~
\ell^-~\bar {\nu _{\ell }}),\\ 
\bar B_a(t) &\equiv & \Gamma (\bar B^0_d(t) \rightarrow  D^{*-}~
\ell^+~ {\nu _{\ell }}),
\end{eqnarray}
where $\bar B_d^0(t) $ denotes the state at time $t$ that evolved from
a state that was pure $\bar B_d^0$ at $t=0$. In the Standard Model the
abnormal rate arises from $B_d\bar B_d$ oscillations. Also, for an
initial $\bar b$ producing a $B_d^0$ , the corresponding normal and
abnormal decay rates are
\begin{eqnarray}
B_n(t) &\equiv & \Gamma (B^0_d(t) \rightarrow  D^{*-}~
\ell^+~{\nu _{\ell } }),\\ 
B_a(t) &\equiv & \Gamma (B^0_d(t) \rightarrow  D^{*+}~
\ell^-~ \bar {\nu _{\ell } }).
\end{eqnarray}
Again $CP$ invariance implies the conditions $B_n(t)=\bar B_n(t)$ and
$B_a(t)=\bar B_a(t)$. 

In what follows, we shall ignore second order $CP$ violations by
neglecting terms that are bilinear in the differences $[B_n(t)-\bar
B_n(t)]$, $[B_a(t)-\bar B_a(t)]$, $(b_n-\bar b_n)$ and $(b_a-\bar
b_a)$.

The number of unmixed events in which a bottom jet is observed on one
side and the particle pair ($D^*\ell $) is present on the other side,
following the primary decay $Z\rightarrow b\bar b$, can be written
(including the $\bar bb$ configuration) as
\begin{eqnarray} 
N_{{\rm unmixed}}
&=& (b_n B_n + b_aB_a)+(\bar b_n\bar B_n +\bar
b_a\bar B_a) \\ 
&\simeq & {1 \over 2}(b_n+\bar b_n)(B_n+\bar B_n )+ {1\over
2}(b_a+\bar b_a)(B_a+\bar B_a ).\label{app1}
\end{eqnarray}
The last step neglects $CP$ violations of second-order. The number of
mixed events which ought to have abnormal charge either at production
or at decay (but not at both), is given by
\begin{eqnarray} 
N_{{\rm mixed}}&=&(b_nB_a+b_aB_n )+(\bar b_n\bar B_a + \bar b_a\bar
                                    B_n)
\\ 
&\simeq & {1\over 2}(b_n+\bar b_n)(B_a+\bar B_a)+ {1\over 2}(b_a+
\bar b_a)(B_n+\bar B_n);\label{app2}
\end{eqnarray}
again, the last step neglects $CP$ violations of second order.  

The observable of interest is the charge-correlation function defined
by
\be 
C_Q(t) = {  N_{{\rm unmixed} }~ -~N_{{\rm mixed}}
           \over N_{{\rm unmixed} }~ +~N_{{\rm mixed}} }~.
\ee
This takes a factorized form when we substitute Eqs. (\ref{app1}) and
(\ref{app2})
\be 
C_Q(t) \simeq K~{ (B_n+\bar B_n) - (B_a+\bar B_a)
\over   (B_n +\bar B_n) + (B_a +\bar B_a)}~.
\ee 
The only assumption behind this simple relation is the neglect of
second and higher order $CP$-violation effects. The time-independent
constant $K$ refers to the jet production tag, while the remaining
factor refers to the decay tag ($K$ is obtained from the decay tag
factor by simply replacing $B_i $ by $b_i$ and $\bar B_i $ by $\bar
b_i$). Hence for convenience in studying the decay time distribution,
we define
\be C'_Q(t) \equiv \frac {1}{K}C_Q(t)~.
\ee

In the framework of the Standard Model, it is reasonable (and
customary) to assume $y\simeq 0$ and that the mass-eigenstates and
$CP$-eigenstates of the neutral beons are nearly the same (which means
$|q/p|\simeq 1$ in the usual notation, $B_{1,2}= pB\pm q\bar B$). We
thus get the standard probabilities per unit time $P_{u,m}(t)$ for
finding, respectively, a $B_d$ or a $\bar B_d $ at time $t$ having
started with an initial $B_d$ \cite{WU},
\begin{eqnarray} 
B_d\rightarrow B_d~:~~ P_u(t)={\Gamma \over 2}e^{-
\Gamma t}(1+\cos \Delta mt),
\\ 
B_d\rightarrow \bar B_d~:~~P_m(t)={\Gamma \over 2}e^{-
\Gamma t}(1-\cos \Delta mt).
\end{eqnarray}
Starting with a $\bar B_d $ at time $t=0$ also leads to the same
unmixed and mixed probabilities per unit time.

For describing the subsequent decays of the beons, we shall appeal to
$CPT$ invariance to relate the semileptonic partial widths of $B_d$
and $\bar B_d$ mesons. If there is a single hadron in the final
semileptonic channel, there is no strong phase due to final state
interactions and the corresponding decay rates of $B_d$ and $\bar B_d$
into individual conjugate channels are equal. Thus the time-dependence
describing the decays into exclusive single-hadron semileptonic
channels will be
\be B_n=\bar B_n = R~P_u(t),\ee 
\be B_a=\bar B_a = R~P_m(t),\ee
with the same constant of proportionality $R$ in the normal and
abnormal cases. In this way we are led to the function
\be C'_Q(t)= \cos \Delta mt \ee
which may be fitted to the observed (proper) time dependence for
extracting the mixing parameter $\Delta m$.

Now we examine how the above analysis gets modified when $\Delta
B=-\Delta Q$ amplitudes are present. Focusing on the specific channel
$D^*(2010)\ell \nu $ , we define the ratios
\be 
\rho =\frac {<D^{*-} \ell ^+\nu _{\ell }|T|\bar B>}
           {<D^{*-} \ell ^+\nu _{\ell }|T|     B>}~,~~~
\bar \rho =\frac {<D^{*+} \ell ^- \bar \nu _{\ell }|T|    B>}
                {<D^{*+} \ell ^- \bar \nu _{\ell }|T|\bar B>}~.
\ee
$CPT$ invariance (the validity of which we assume throughout) implies
the relation $\bar \rho =\rho ^*$ . On the other hand, $CP$ invariance
implies $\bar \rho =\rho $. However when we allow for the presence of
$\Delta B=-\Delta Q$ amplitudes, it is reasonable to allow also for
possible $CP$ violations. We do this by retaining the terms which are
linear in the $CP$ violating quantities $(\rho -\bar \rho )$ and
$(|p|^2-|q|^2)$ (where $p$ and $q$ are the coefficients which define
the propagation states as mixtures of flavour states).

We reevaluate the charge correlation function by neglecting terms
that are second (and higher) order small in $\rho $ or/and $CP$
violations, to obtain
\be 
C'_Q(t)\simeq \frac {\cos \Delta mt }{\cosh y\Gamma t~+~ 2 {
\rm Re}~ \rho~ \sinh y \Gamma t }.
\ee    

Instead of fitting the data to the above function, we consider the
corresponding time-integrated version
\be 
{\cal C'} = \frac {a} {1+ 2y~ {\rm Re}~ \rho }~~,~~a\equiv \frac
{1-y^2}{1+x^2}~,
\ee 
and obtain information on $y{\rm Re}~\rho $. For the parameter $a$ we
use the relation connecting it to the like-sign dilepton fraction
$\chi _d$ \cite{OK},
\be \chi _d= \frac {1-a}{2}.\ee
From the experiments at $\Upsilon (4S)$ by ARGUS \cite{ARG94} and CLEO
\cite{CLE93} collaborations, we have the average value \cite{PDG}
\be \chi _d= 0.156 \pm 0.024~. \ee 
By definition $\chi _d$ is $CP$ even and hence unaffected by first
order $CP$ violations; it is also unaffected by terms which are linear
in $\rho $ \cite{DS}. Therefore, substituting for $a$, we obtain
\be {\cal C'} = \frac {1-2 \chi _d} {1+ 2y~ {\rm Re}~ \rho}~.\ee
A determination of the combination $y {\rm Re}~\rho $ is thus possible
provided we know ${\cal C'}$. We observe that the experimental value
for $x_d$ deduced from $B\bar B$ oscillations at $Z$ does give
us ${\cal C'}$ through the relation
\be {\cal C'}={1\over (1+x_d^2)}~\ee
because $y=0$ and $\rho =0 $ are assumed in the experimental
determinations of $x_d$.
   
We consider, mainly to illustrate our procedure, the oscillation data
in $Z$ decays using the $D^*\ell / Q_J$ method. This method uses the
jet charge $Q_J$ for production tag and
hence deals with a bigger event sample than that using lepton tag.
Also, it uses for the decay tag, the semi-exclusive channel $
B_d\rightarrow D^*(2010)\ell \nu X $ for which contamination from the
semileptonic decays of $B_s$ and charged $B$ is expected to be
minimal. Recently the OPAL group \cite{OP96} had reported a value for
$\Delta m$ using this method; it is based on a sample of 1200 $D^{*\pm
}\ell ^{\mp } $ candidate events of which 778 $\pm $84 are expected to
be from $B_d^0$ decays. Multiplying this $\Delta m$ by the average
lifetime of the $B_d$ meson \cite{PDG}, we obtain
\begin{eqnarray} 
x_d &=&(0.539 \pm 0.060 \pm 0.024)~{\rm ps}^{-1}~.~(1.56 
                                          \pm 0.06)~{\rm ps}~\\ 
    &=& 0.84\pm 0.11.
\end{eqnarray}
Assuming that the observed events are all due to the 3-body channel
$B_d\rightarrow D^*(2010)\ell \nu $ , we are therefore led to conclude
that
\begin{eqnarray} 
y {\rm Re}~\rho &=& {1\over 2}[(1-2\chi _d)(1+x_d^2)-1]\\
                &=& 0.09\pm 0.07.
\end{eqnarray}
A more accurate determination of this and other such quantities should
be possible \cite{DS96} with the future dilepton data at the
asymmetric $B$ Factories.

To conclude, we have shown that the magnitude of the product $y {\rm
Re}~\rho $ cannot exceed 0.21 at 90\% CL. This limit depends on the
dilepton data at $\Upsilon (4S)$ and the $B_d\bar B_d$ oscillation
data from $Z$ decays in which the decaying neutral beon is tagged by
the pair $D^*(2010)^{\mp }\ell ^{\pm }$.

\end{document}